\documentclass[12pt,a4paper,twoside]{article}
\usepackage{graphicx}
\usepackage{baltlat1}
\begin{document}
\ \
\vspace{0.5mm}

\setcounter{page}{1}
\vspace{8mm}

\titlehead{Baltic Astronomy, vol.~12, XXX--XXX, 2003.}

\titleb{ON THE FEASIBILITY OF DISK CHEMICAL\\ MODELING}

\begin{authorl}
\authorb{D.~Semenov}{1},
\authorb{Y.~Pavluchenkov}{2},
\authorb{Th.~Henning}{3},
\authorb{E.~Herbst}{4}, and\\
\authorb{E.~van Dishoeck}{5}
\end{authorl}

\begin{addressl}
\addressb{1}{Astrophysical Institute and University Observatory,
Schillerg{\"a}{\ss}chen 3, 07745 Jena, Germany}

\addressb{2}{Insitute for Astronomy of the Russian Academy of Sciences,\\
Pyatnitskaya ul. 48, 119017 Moscow, Russia}

\addressb{3}{Max Planck Insitute for Astronomy,
K{\"o}nigstuhl 17, 69117 Heidelberg, Germany}

\addressb{4}{Departments of Physics, Chemistry, and Astronomy, 
Ohio State\\ University, 140 West 18th Avenue, Columbus, OH 43210-1173}

\addressb{5}{Sterrewacht Leiden, 
PO Box 9513, 2300 RA Leiden, The Netherlands}

\end{addressl}

\submitb{Received October 30, 2003}

\begin{abstract}
In this paper, we compare the results of the modeling of a protoplanetary disk chemical evolution
obtained with the UMIST\,95 and ``New Standard Model'' (NSM) chemical databases. Assuming the same
initial conditions, it is found that the substitution of one chemical network by another causes 
almost no difference for the disk ionization degree. In contrast, the NSM and UMIST\,95 
abundances of CO can differ by a factor of a hundred at some regions of the disk surface. However, 
relevant CO vertical column densities differ much less, at most by a factor of a few. In addition, 
we synthesize the single-dish CO(J=3-2) line by means of the 2D line radiative transfer for both 
considered chemical networks. It is shown that the intensity of this line in the case of the UMIST\,95 
abundances is lower compared to the NSM case by $\sim 15\%$.
\end{abstract}


\resthead{On the feasibility of disk chemical modeling}{D.~Semenov et al.}

\sectionb{1}{INTRODUCTION}
Nowadays computational facilities allow to perform extensive simulations of the chemical evolution of 
various cosmic objects, like (collapsing) protostellar clouds or protoplanetary disks (e.g., Li et al.~2002;
Aikawa et al.~2002). In these studies usually either the UMIST\,95 (Millar et al.~1997) or ``New Standard Model'' 
(Aikawa \& Herbst~1999) database of chemical reactions is utilyzed. As the result of the modeling, 
time-dependent abundances of chemical species as well as their column densities can be obtained 
and faced with observational data. However, due to the lack of the full information about the physical conditions 
in the object under investigation and/or the complexity of the feasible modeling itself, typically these values 
are calculated with a certain (unknown) degree of accuracy. Therefore, it would be very useful to estimate, even
approximately, how high it could be. The primary goal of this study is to verify how sensitive are the computed 
molecular abundances in the case of a protoplanetary disk chemistry to the adopted set of chemical reactions.
Second, if such a difference exists for a certain species, we find by means of the reduction technique
which chemical reactions are responsible for that. Finally, we study how such an abundance 
difference may affect the results of line radiative transfer calculations using the 2D line radiative
transfer code ``URAN(IA)''.
\vskip4mm

\sectionb{2}{DISK MODEL}

\subsectionb{2.1}{Disk physics}
We used essentially the same disk physical and chemical model as described in Semenov et al.~(2004) (herefater Paper~I).
Briefly, the 1+1D flared steady-state accretion disk model of D'Alessio et al.~(1999) is adopted with 
a radius of 373~AU, accretion rate $\dot{M}=10^{-7}\,M_\odot$~yr$^{-1}$, and viscosity parameter 
$\alpha=0.01$. The central star is assumed to be a classical T Tau star with an effective temperature 
$T_*=4\,000$~K, mass $M_*=0.5M_{\odot}$, and radius $R_*=2R_{\odot}$. We took into account the illumination
of the disk by the stellar UV radiation with the intensity $G_*=10^4\,G_0$ at 100~AU, where $G_0$ is
the intensity of the mean interstellar UV field by Draine~(1978) as well as interstellar UV radiation,
but neglected stellar X-rays. The cosmic rays are considered as another ionizing source with 
$\zeta_0=1.3\cdot10^{-17}$~s$^{-1}$ unattenuated ionization rate. To calculate the propogation of the UV radiation 
and cosmic rays through the disk, Eqs.~(2)--(3) from the Paper~I are used. We did not take into account ionization by 
the decay of radionucleides.

\subsectionb{2.2}{Chemical model}
Contrary to Paper~I, a gas-grain chemistry without surface reactions is considered. In addition to our model which is
based on the UMIST\,95 ratefile, we adopted the NSM network (Herbst, priv. communication). The latter was extended by
a set of reactions of dissociative recombinations on neutral and negatively charged grains as well as grain charging 
processes. We added a reaction of the H$_2$ photodissociation from the NSM model to our UMIST\,95 network with 
unattenuated rate $k_0=3.4\cdot10^{-11}$~cm$^3$~s$^{-1}$. All the rest parameters 
of the chemical model from Semenov et al.~(2004) were the same. As the initial abundances, we used the same set of the 
``low metals'' abundances from Wiebe et al.~(2003) for the entire disk.

\sectionb{3}{RESULTS}
\begin{figure}
\centering
  \includegraphics[width=\textwidth,clip=]{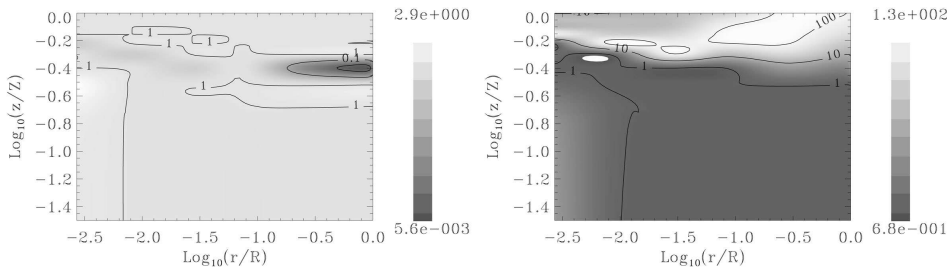}
  \caption{Shown are the distributions of the NSM-to-UMIST\,95 abundance ratios calculated
  for e$^-$ {\bf (left panel)} and CO {\bf (right panel)} over the disk.}
\label{abunds}
\end{figure}

\begin{figure}
\centering
  \includegraphics[width=\textwidth,clip=]{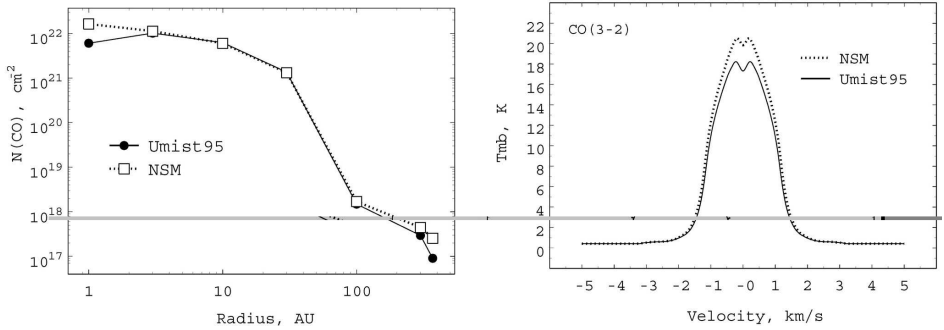}
  \caption{The radial distribution of the CO vertical column densities is presented
  on the {\bf left panel}. The solid line with filled circles corresponds to the case of the UMIST\,95 
  database, while the dashed line with open squares is for the NSM chemical network. On the {\bf right
  panel}, the synthetic single-dish CO(3-2) line profiles calculated with the UMIST\,95
  (solid line) and NSM (dotted line) abundances of CO are depicted. It is assumed that the disk 
  has the distance of 150~pc, its inclination angle is $30^\circ$, and the beam size of the telescope 
  is $10^{\prime\prime}$.}
\label{coldens}
\end{figure}
With the disk physical model and two chemical networks described in the previous section, we modeled the disk chemistry 
for 1~Myr evolutionary time span. The 2D disk distributions of the NSM-to-UMIST\,95 ratio of the resulting fractional 
ionization and CO abundances are plotted in Figure~1, left and right panels, respectively. As clearly seen,
there is no difference for disk ionization degree in the entire disk but one small region, $r\sim350$~AU, $z\sim235$~AU,
where it can reach a factor of 200. On the constrary, 
$n^\mathrm{NSM}(\mathrm{CO})/n^\mathrm{UMIST}(\mathrm{CO})>1$ and can be as high as about 100 in the disk surface,
$\log(z/Z)\sim-0.4$. However, the maximum CO gas-phase concentration is reached in the disk intermediate layer,
$-1.0<\log(z/Z)<-0.3$, where both the UMIST\,95 and NSM networks give approximately the same values of the carbon monoxide
abundances. Therefore, the corresponding CO vertical column densities do not deviate much, at most by a factor of several
for the disk inner and outer radii (see Figure~2, left panel). Using the UMIST\,95 and NSM disk abundances of CO, we 
synthesized the single-dish CO(J=3-2) spectra with the 2D line radiative transfer code ``URAN(IA)'' (Pavluchenkov et 
al.~2003), which are plotted in Figure~2, right panel. It can be clearly seen that the intensities of this line differ 
by about 15\%, which is lower than typical 30\%--50\% uncertainty of the observational data. 

To figure out what is the reason for the difference in CO disk abundances, we used the reduction technique from 
Wiebe et al.~(2003) and isolated two small subsets of primary carbon monoxide formation and destruction routes 
for both chemical networks. Surprisingly, it is found that these two sets contain nearly the same chemical reactions
but some of them differ by the adopted rate coefficient values. Among the reactions with different rates, we designated
three most important ones, which are listed in Table~1. Note that the corresponding rate values are given for the disk
location $r=10$~AU, $z=3.15$~AU, where gas temperature and particle density are $T=143$~K and $n=3.3\cdot10^{7}$~cm$^{-3}$, 
the UV intensity and visual extinction are $G_*=9.1\cdot10^5$ and $A_\mathrm{V}=1.2^\mathrm{m}$, and 
ionization rate is $\zeta=7.45\cdot10^{-18}$~s$^{-1}$.

\begin{center}
\vbox{\footnotesize
\begin{tabular}{l@{\hskip10mm}r@{\hskip10mm}rr}
\multicolumn{3}{c}{\parbox{120mm}{\baselineskip=9pt
{\smallbf\ \ Table 1.}{\small\ Chemical reactions which cause the most of the CO disk abundance difference.}}}\\
\tablerule
Reaction  & UMIST\,95 rate,~cm$^3$~s$^{-1}$  & NSM rate,~cm$^3$~s$^{-1}$\\
\tablerule
C$^+$ + OH $\rightarrow$ CO$^+$ + H    &  $7.7\cdot10^{-10}$  & $8.3\cdot10^{-9}$  \\
CO$^+$ + H $\rightarrow$ CO + H$^+$    &  $7.5\cdot10^{-10}$  & $9.0\cdot10^{-11}$  \\
CO + $h\nu$ $\rightarrow$ C + O        &  $9.1\cdot10^{-6}$   & $1.4\cdot10^{-6}$  \\
\tablerule
\end{tabular}
}
\end{center}
 
\sectionb{4}{SUMMARY}
The sensitivity of the results of the disk chemical modeling to the applied chemical databases
is investigated. It is found that the modeled disk fractional ionizations are nearly identical
for both the UMIST\,95 and NSM networks, while the CO abundances can vary by a factor of
a hundred in some parts of the disk. However, the corresponding vertical column densities of CO do 
not show such a strong variation with the adopted set of chemical reactions. Consequently, it 
results in a small ($\sim 15\%$) difference between the synthetic single-dish CO(3-2) line intensities
obtained with two considered chemical sets.

ACKNOWLEDGMENTS.\ DS was supported by the {\it Deutsche 
Forschungsgemeinschaft}, DFG project "Research Group Laboratory 
Astrophysics'' (He 1935/17-2), the work of YP was supported by the 
RFBR grants 01-02-16206 and 02-00-04008.
\goodbreak

\References
\ref Li~Z.--Y., Shematovich~V.~I., Wiebe~D.~S., Shustov~B.~M. 2002, ApJ, 569, 792

\ref Aikawa~Y., van Zadelhoff~G.--J., van Dishoeck~E.~F., Herbst~E. 2002, A\&A, 
386, 622

\ref Millar~T.~J., Farquhar~P.~R.~A., Willacy~K. 1997, A\&AS, 121, 139

\ref Aikawa~Y., Herbst~E. 1999, A\&A, 351, 233

\ref Wiebe~D.~S., Semenov~D.~A., Henning~Th. 2003, A\&A, 399, 197

\ref Semenov~D.~A., Wiebe~D.~S., Henning~Th. 2004, A\&A, submitted

\ref D'Alessio~P., Calvet~N., Hartmann~L. et al. 1999, ApJ, 527, 893

\ref Draine~B.~T. 1978, ApJS, 36, 595 

\ref Pavluchenkov~Y., Shustov~B.~M. et al. 2003, Astronomy Reports, in prep.

\end{document}